\documentclass[a4paper,11pt]{article}
\usepackage{pos}

\newcommand{\Emax}{E_\text{max}}
\newcommand{\Emaxl}{E_\text{max,l}}
\newcommand{\Emaxh}{E_\text{max,h}}
\newcommand{\alphal}{\alpha_\text{l}}
\newcommand{\alphah}{\alpha_\text{h}}
\newcommand{\Lc}{L_\text{c}}
\newcommand{\dtFlare}{\Delta t_\text{flare}}
\newcommand{\dtAgn}{\Delta t_\text{AGN}}
\newcommand{\dtIgmf}{\Delta t_\text{IGMF}}
\newcommand{\txs}{TXS~0506+056}

\title{Multimessenger Constraints on Intergalactic Magnetic Fields from Flaring Objects}
 \ShortTitle{Multimessenger Constraints on Intergalactic Magnetic Fields from Flaring Objects}

\author*[a,b]{Andrey Saveliev}
\author[c]{{Rafael}~{Alves Batista}}

\affiliation[a]{I. Kant Baltic Federal University, Institute of Physics, Mathematics and Information Technology, \\
ul. Nevskogo 14 A, 236016 Kaliningrad, Russia}

\affiliation[b]{Lomonosov Moscow State University, Faculty of Computational Mathematics and Cybernetics,\\ 
Leninskiye Gory 1-52, 119991 Moscow, Russia}

\affiliation[c]{Radboud University Nijmegen, Department of Astrophysics/IMAPP,\\ 
P.O. Box 9010, 6500 GL Nijmegen, The Netherlands}

\emailAdd{andrey.saveliev@desy.de}
\emailAdd{r.batista@astro.ru.nl}

\abstract{The origin of magnetic fields in the Universe is an open problem. Seed magnetic fields possibly produced in early times may have survived up to the present day close to their original form, providing an untapped window to the primeval Universe. The recent observations of high-energy neutrinos from the blazar TXS 0506+056 in association with an electromagnetic counterpart in a broad range of wavelengths can be used to probe intrinsic properties of this object and the traversed medium. Here we show that intergalactic magnetic fields (IGMFs) can affect the intrinsic spectral properties of this object reconstructed from observations. In particular, we point out that the reconstructed maximum gamma-ray energy of TXS 0506+056 can be significantly higher if IGMFs are strong. Finally, we use this flare to constrain both the magnetic-field strength and the coherence length of IGMFs.}

\FullConference{37$^{\rm{th}}$ International Cosmic Ray Conference (ICRC 2021)\\
		July 12th -- 23rd, 2021\\
		Online -- Berlin, Germany}

\begin{document}
\maketitle

\section{Introduction}

Up to the present day little is known about Intergalactic Magnetic Fields (IGMFs), i.e.~magnetic fields in the voids of the intergalactic space. This lack of knowledge starts with their origin, as it is not certain whether IGMFs were created through cosmological events in the very early Universe (like Inflation or a cosmological phase transition) or during structure formation due to astrophysical processes like primordial vorticity, galactic outflows, and so on (for a review of different magnetogenesis scenarios see, for example, \cite{kulsrud2008a,durrer2013a,vachaspati2021a}). However, also the measurement of IGMFs poses a challenging task -- so far, most of the used techniques are only able to infer {\it upper} limits on the magnetic field strength, based on Faraday rotation and Zeeman splitting measurements (as described in \cite{neronov2009a}), cosmological observations, especially of the cosmic microwave background (CMB) -- see \cite{jedamzik2019a} for a concise overview -- and the detection of gamma rays from blazars (see \cite{alvesbatista2021a} for an extensive review of the status of the field). Especially the latter method has sparked some interest due to the fact that the authors of \cite{neronov2010a} claimed to be able to set a {\it lower} limit on the magnetic field strength, even though this claim has been disputed as other authors claim that instead also other effects, in particular plasma instabilities, might explain the observations \cite{broderick2018a,alvesbatista2019a}.

In this work we extend the idea of using gamma ray observations to determine the properties of IGMFs by discussing a novel method which, in addition to gamma rays, takes neutrinos from individual flaring objects into account. We illustrate the method using the blazar \txs. In particular, we derive limits on the correlation length of the magnetic field \cite{alvesbatista2020a}, and show the importance of taking into account IGMFs when determining intrinsic properties of gamma-ray sources \cite{saveliev2021a}.

We structure this article as follows: In Sec.~\ref{sec:ModelsSims} we present the models of the source and of the gamma-ray propagation together with the corresponding simulation setup. In Sec.~\ref{sec:DataAnalysis} we show our data analysis, in particular the various fits of our simulations to the observations of \txs. In Sec.~\ref{sec:Discussion} we discuss our results before summarizing our work and giving a short outlook in Sec.~\ref{sec:SummaryOutlook}.

\section{Models and Simulations} \label{sec:ModelsSims}

The observation which is the basis of this work is the detection of the high-energy neutrino event IC-170922A by the IceCube Observatory~\cite{icecube2018a,icecube2018b} as well as of its electromagnetic counterpart at different wavelengths~\cite{icecube2018b}. These detections have been associated with a flare of the blazar \txs, located at $z \simeq 0.3365 \pm 0.0010$~\cite{paiano2018a}. The two periods of increased activity which are considered here were measured by the Major Atmospheric Gamma-ray Imaging Cherenkov (MAGIC) telescope were observed on MJD 58029.22 and MJD 58030.24, at $E > 9 \times 10^{10} \; {\rm eV}$~\cite{magic2018a}. The hypothesis that the correlation between the electromagnetic and the neutrino signals happened by chance is rejected at a  $3\sigma$-level~\cite{icecube2018b}. 

The spectral parameters of the period with enhanced emission (denoted as ``high'') are not necessarily the same as the ones during the normal state (denoted as ``low''). Thus, we model the spectrum of gamma rays effectively emitted by the source as
\begin{equation}
\frac{dN}{dE} = J_{0}
\begin{cases}
E^{-\alphal} \exp\left( -\frac{E}{\Emaxl} \right) & \text{for the low state}\,, \\
\eta E^{-\alphah} \exp\left( -\frac{E}{\Emaxh} \right) & \text{for the high state}\,,
\end{cases}
\end{equation}
where $J_{0}$ is an overall normalization factor, $\eta$ denotes the flux enhancement in the high state (subscript `h') with respect to the low state (`l'), while $\alpha_{i}$ and $E_{\text{max},i}$ are the corresponding spectral indices and maximum energies. This is computed within a time interval $\dtFlare$ which is given by the duration of the neutrino flare. The second relevant time scale is the one over which {\txs} is a gamma-ray emitter at the low state, denoted by $\dtAgn$. In general, the times of AGN activity range from~$10^{6}$ to $10^{8}$ years~\cite{parma2002a}. We use $\dtAgn = 10$, $10^{4}$, and $10^{7}$ years. 

While neutrinos propagate almost without interacting, photons are subject to electromagnetic interactions: They initiate an electromagnetic cascade by interacting with photons of the CMB and the extragalactic background light (EBL), resulting in the creation of electron-positron pairs: $\gamma + \gamma_{\rm bg} \rightarrow e^{+} + e^{-}$, where $\gamma$ is the gamma ray and $\gamma_{\rm bg}$ the background photon, respectively. These electrons and positrons subsequently react with the background radiation, upscattering the CMB/EBL photons to high energies via  inverse Compton ($e^\pm + \gamma_{\rm bg} \rightarrow e^\pm + \gamma$). The high-energy photons produced will then restart the whole process, creating a cascade of particles. This cascade stops once the energy of the photons drops below the kinematic threshold for pair production. Therefore, the gamma-ray signal detected at Earth consists both of primary photons arriving directly from the source, without interacting during propagation, and of secondary photons resulting from the cascading process described above.

The electrons and positrons of the cascade are sensitive to the local properties of the magnetic field, as they are deflected in opposite directions due to the Lorentz force. The magnetic field itself is characterized by a Kolmogorov spectrum, the magnetic field strength $B$ and the coherence length $\Lc$.

We simulate the development of electromagnetic cascades in the magnetized intergalactic space using the CRPropa~3 code~\cite{alvesbatista2016a}, considering pair production, inverse Compton scattering, adiabatic losses due to the expansion of the Universe, and synchrotron emission. We simulated various scenarios by scanning over the parameters $B$, $\Lc$, the spectral index $\alpha$ and the cutoff energy $\Emax$ of the source spectrum (for which the values of the latter two may be used for the low and high states -- see above) in the ranges $10^{-19} \leq B / \text{G} \leq 10^{-14}$ (and, in addition, the case $B=0$), $10^{-2} \leq \Lc / \text{Mpc} \leq 10^{3}$, $0 \leq \alpha \leq 4$, and $10^{10} \leq \Emax/ {\rm eV} \leq 10^{14}$, respectively. We use the following EBL models: \cite{gilmore2012a,dominguez2011a}, and the lower and uopper limits from~\cite{stecker2016a}.

\begin{figure}
	\centering
	\includegraphics[width=0.96\textwidth]{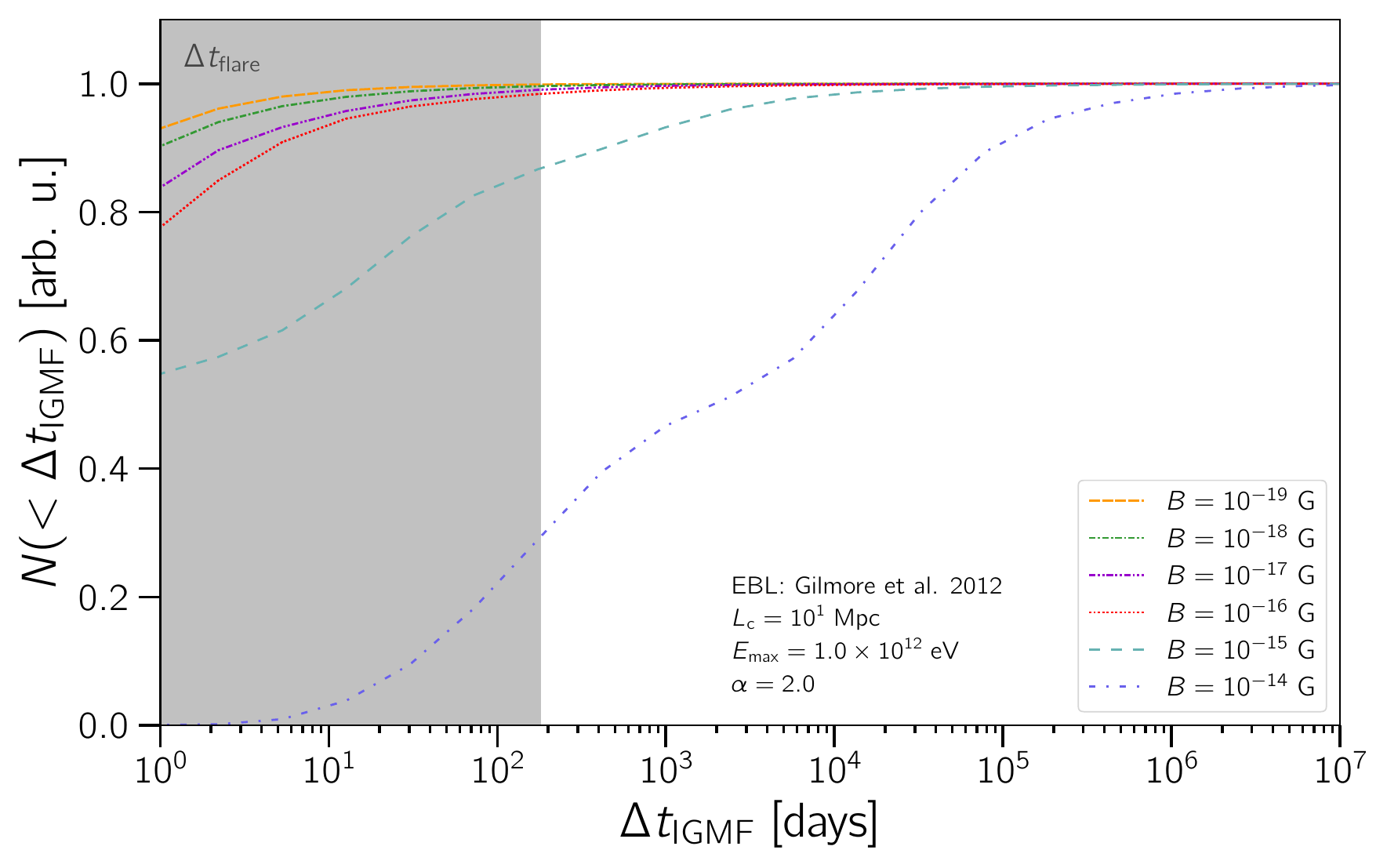}
	\caption{Cumulative distribution of time delays of the gamma rays ($\dtIgmf$). The grey shaded region indicate the period of enhanced activity of the object ($\dtFlare$). The scenario shown here is for $\Emax=1 \; \text{TeV}$ and $\alpha=2$, assuming a coherence length $\Lc=10 \; \text{Mpc}$.}
	\label{fig:dt}
\end{figure}

Our constraints are based on the time delay ($\dtIgmf$) of the arriving gamma-ray signal with respect to the neutrino flaring period ($\dtFlare$). To understand this effect, in Fig.~\ref{fig:dt} we show the cumulative distribution of time delays for one specific intrinsic spectrum of the object and different magnetic field strengths. Note that for stronger magnetic fields ($B \gtrsim 10^{-15} \; \text{G}$) more than $10\%$ of the gamma rays do not arrive at Earth within a time window of $\dtFlare=180 \; \text{days}$, corresponding to the neutrino flare~\cite{icecube2018a, icecube2018b}. This is the essence of the multimessenger method we present here.

\section{Data Analysis} \label{sec:DataAnalysis}

\begin{figure}[htb]
	\centering
	\includegraphics[width=0.96\textwidth]{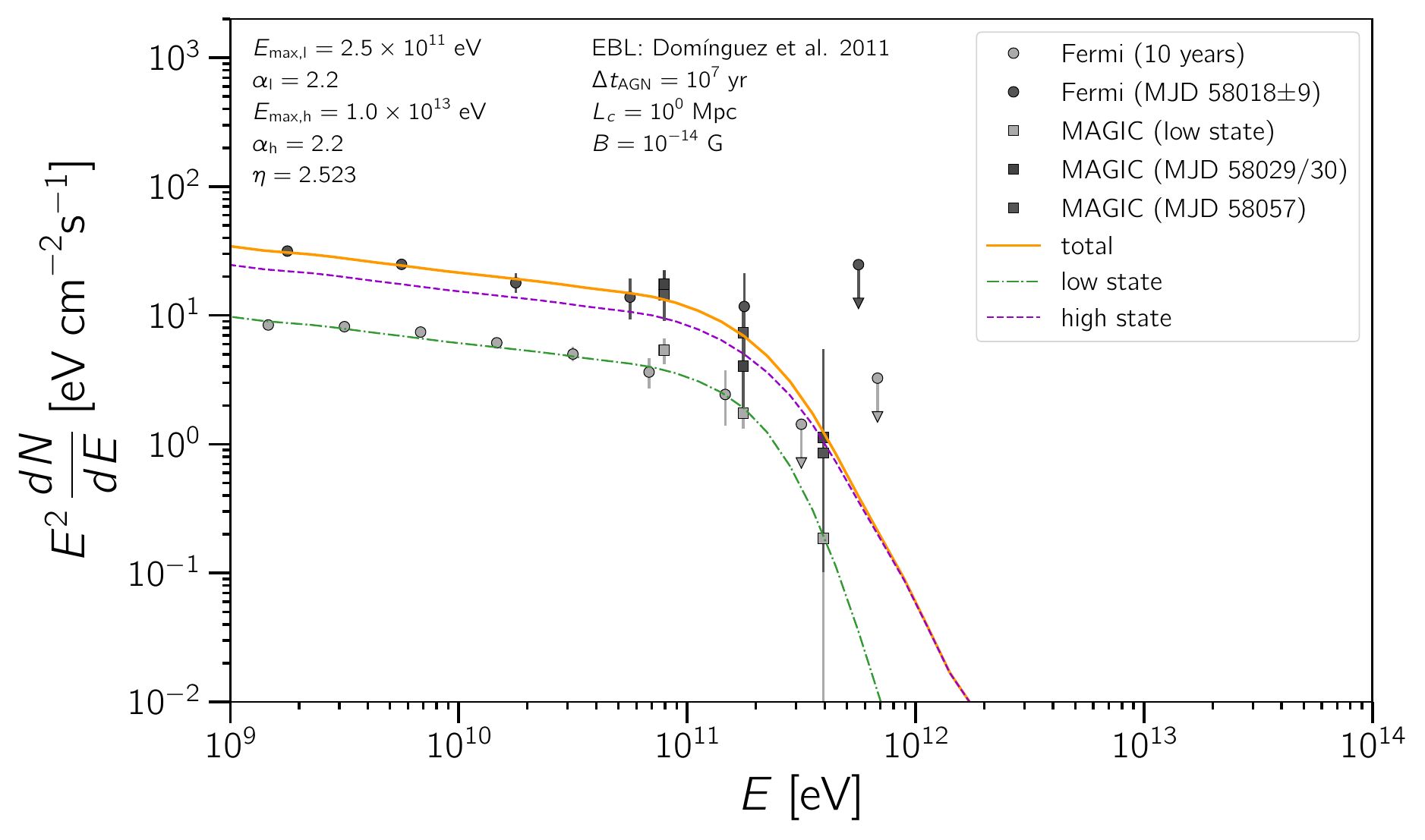}
	\caption{An example of the fit to the observed flux. The dot-dashed green line represents the fit to the low state. The dashed purple line corresponds to the high state ignoring the contribution of the low state. The combined total (low$+$high state) is shown as a thick orange solid line.}
	\label{fig:spec}
\end{figure}

We are now able to constrain IGMFs using information from both messengers -- gamma rays and neutrinos. First, we fit the spectrum for the low state with respect to the magnetic field parameters and find that the spectral parameters remain virtually unaltered regardless of the magnetic-field properties: $\alphal=2.2$ and $\Emaxl = 250 \; \text{GeV}$ if we only consider combinations of $\alpha_\text{l}$ and $\Emaxl$ for which the fit produces $p$-values $p > 10^{-3}$. The second step is to use these values to scan over all the combinations of the parameters $\Emaxh$, $\alphah$, $B$, and $\Lc$. One example of the fitted spectrum is shown in Fig.~\ref{fig:spec}.

To show that the IGMF is non-zero, we carried out a hypothesis test for each of the four considered EBL models, with the null hypothesis being $B=0$. To do so, we marginalized over all other quantities, obtaining a probability distribution for the simulated values of $B$. We found that only for the models in~\cite{dominguez2011a} and the lower limit model in~\cite{stecker2016a} the null hypothesis can be rejected. For completeness, we also include the EBL models from~\cite{gilmore2012a} and the upper limit from~\cite{stecker2016a} in our considerations 
for which we cannot reject the hypothesis of $B=0$, but that still allows for non-zero magnetic fields. This should be kept in mind in what follows.

In order to constrain the magnetic field $B$ and the coherence length $\Lc$, we first marginalize our results over the spectral parameters for different values of $\dtAgn$ (as it turns out, $\dtAgn$ has a small impact on the constraints). We then derive two-dimensional marginalized confidence regions for $B$ and $\Lc$, as shown in Fig.~\ref{fig:fitB}.

\section{Discussion} \label{sec:Discussion}

\begin{figure}[hbt]
	\centering
	\includegraphics[width=.49\textwidth]{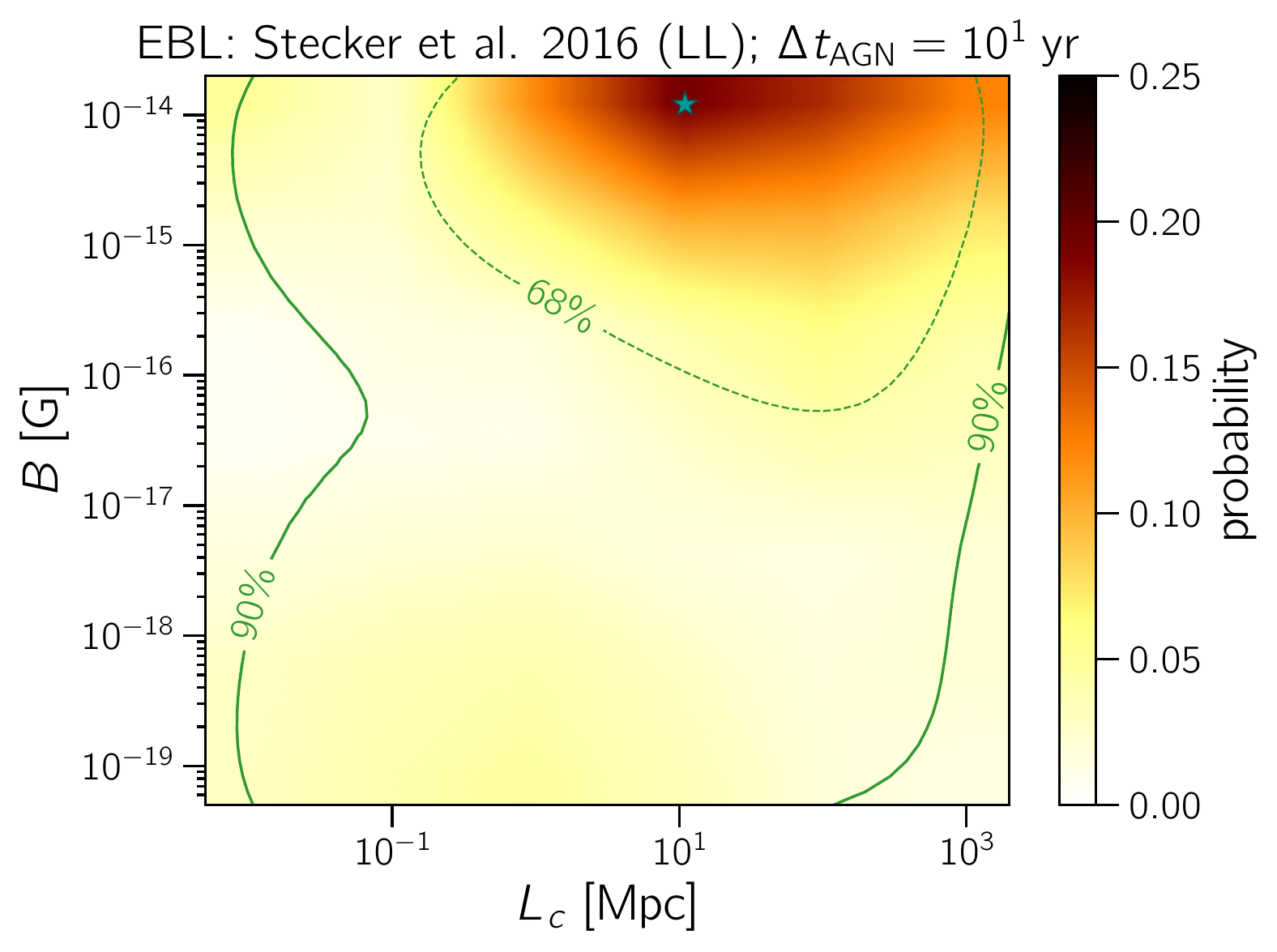}
	\includegraphics[width=.49\textwidth]{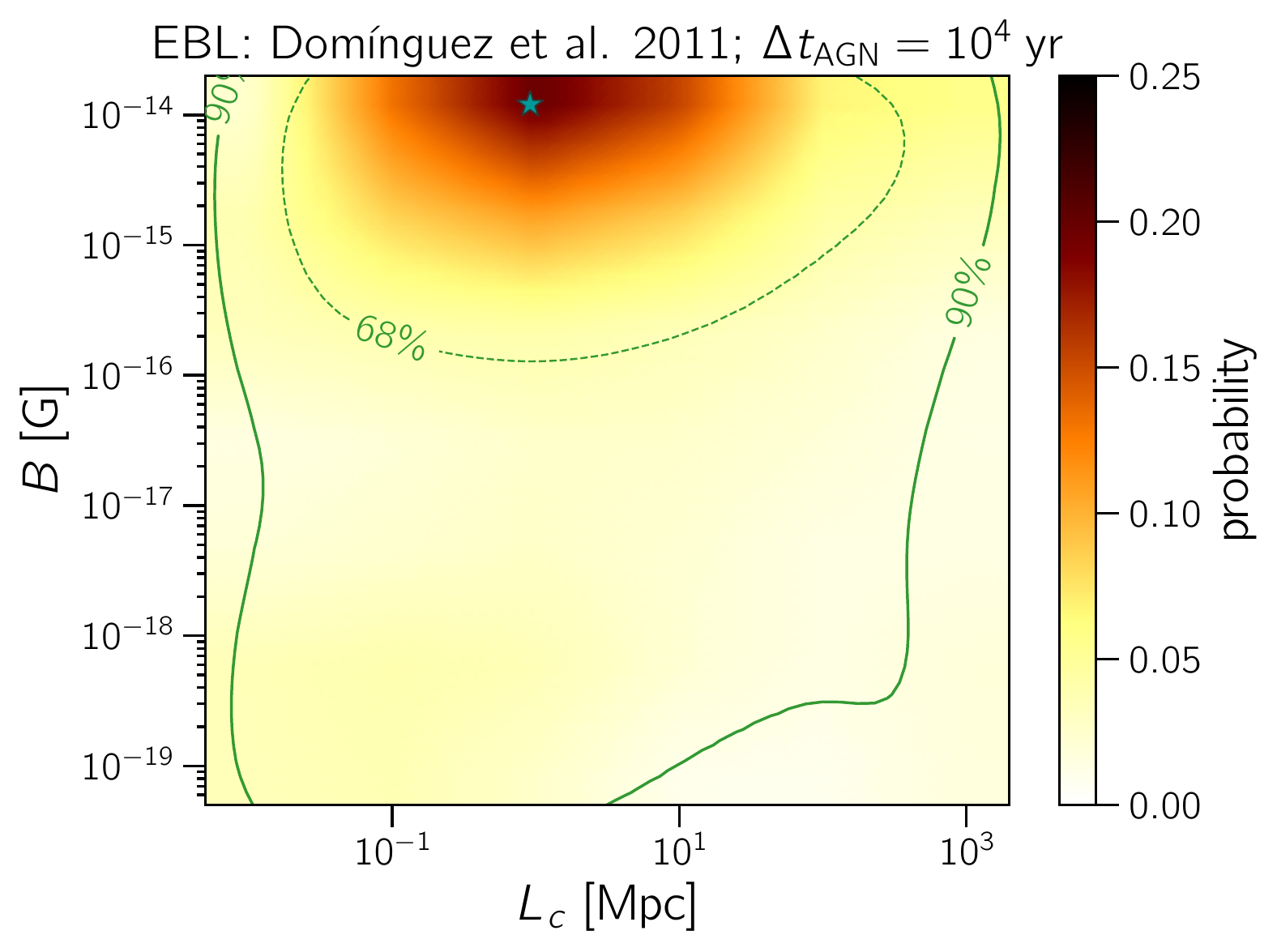}
	\caption{Results of the fit marginalized over the spectral parameters ($\Emax$ and $\alpha$) for the EBL models and $\dtAgn$ indicated in the figures. The colour scale denotes the probability normalized to unit. The star indicates the best-fit point.}
	\label{fig:fitB}
\end{figure}

Our results shown in Fig.~\ref{fig:fitB} provide seemingly weak constraints on the parameter space. For example, with the EBL from~\cite{dominguez2011a}, the 90\% contour disfavors only very large coherence lengths ($\Lc \gtrsim 100 \; \text{Mpc}$) for fields weaker than $B \sim 10^{-18} \; \text{G}$. If IGMFs have galactic scales ($\Lc \lesssim 10-100 \; \text{kpc}$), then $B \gtrsim 10^{-18} \; \text{G}$ (at 90\% C.L.). Similarly, for the lower-limit model from~\cite{stecker2016a} , $\Lc \lesssim 10 \; \text{kpc}$ is not contained within the 95\% confidence region.

Coherence lengths of $\sim 10 \; \text{kpc}$ are rejected at a 90\% confidence level for the two EBL models shown in Fig.~\ref{fig:fitB}. This would disfavor models in which the intergalactic space is magnetized by galactic winds (see, e.g.,~\cite{bertone2006a}), since they predict $\Lc \sim 1-10 \; \text{kpc}$. Models in which IGMFs were generated by cosmic rays escaping from galaxies prior to Reionization (see, for example,~\cite{miniati2011a}) are only slightly compatible with our results. On the other hand, fields originating from AGNs are thought to have $\sim$Mpc scales~\cite{durrer2013a}, and therefore lie well within the estimated limits. Within the scope of this work, we did not test $\Lc \lesssim 10 \; \text{kpc}$, which, however, is important, as it would allow us to constrain certain models of cosmological magnetogenesis \cite{durrer2013a}.

\begin{figure}
  \includegraphics[width=0.49\textwidth]{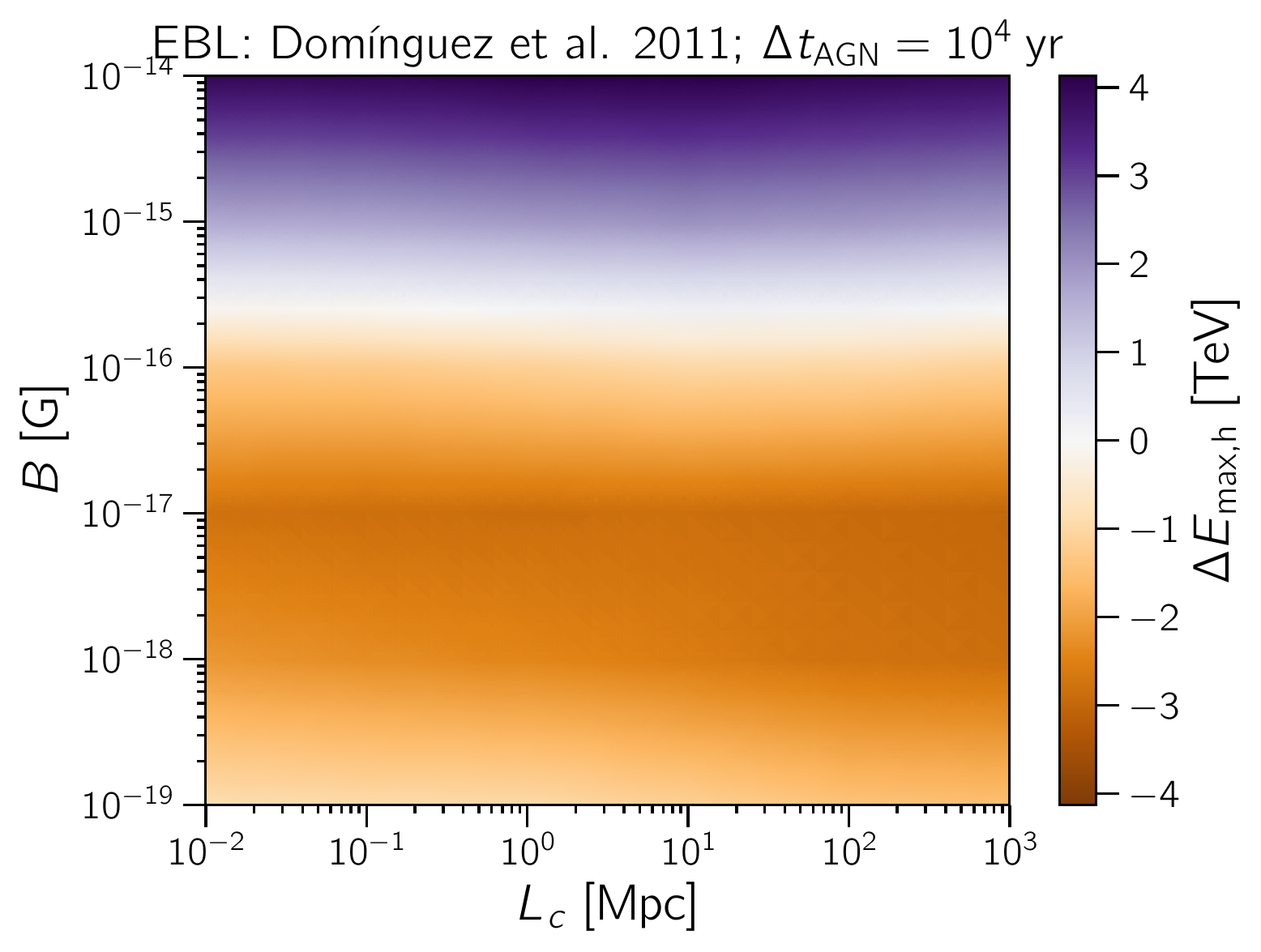}
  \includegraphics[width=0.49\textwidth]{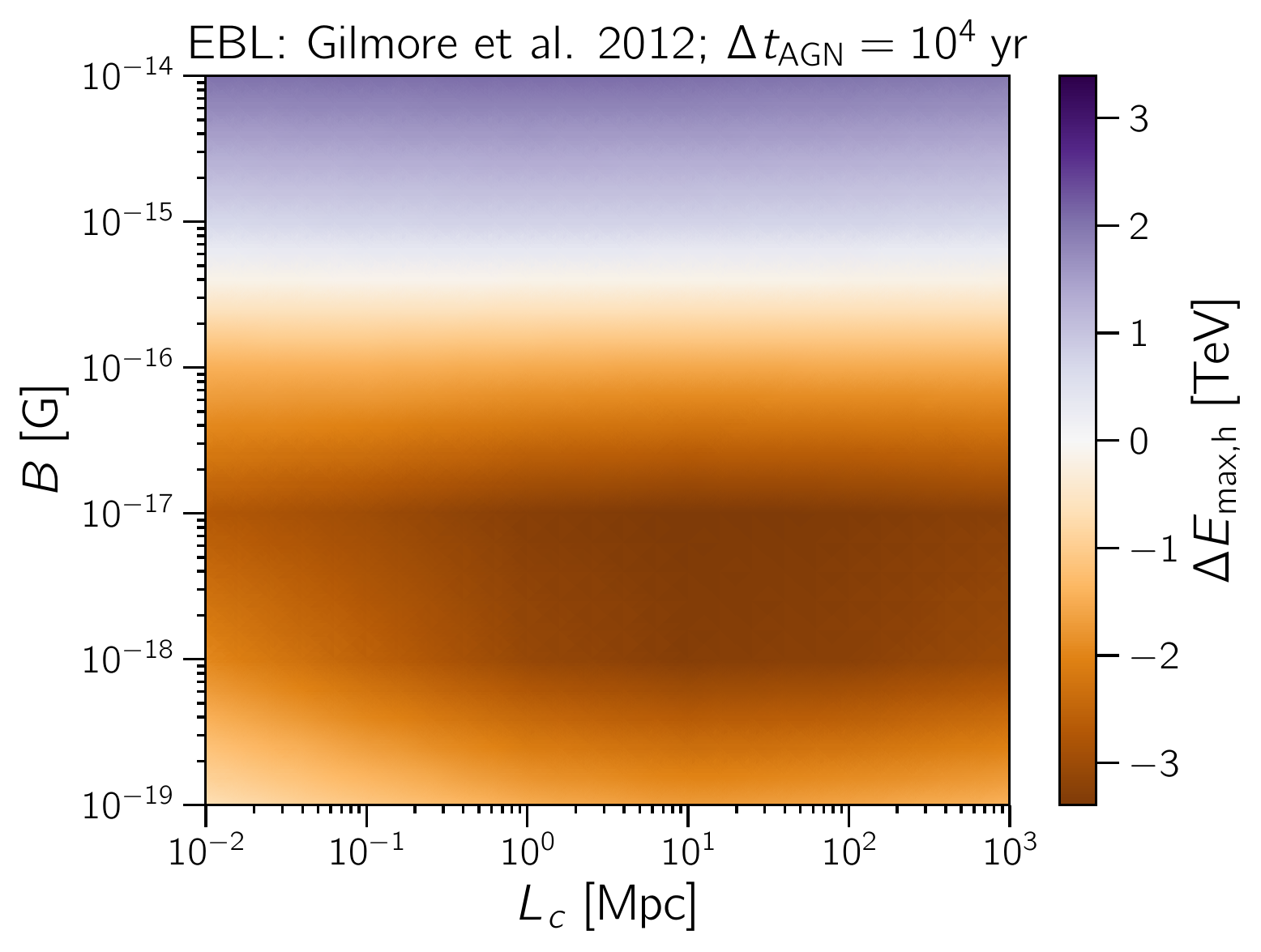}
  \includegraphics[width=0.49\textwidth]{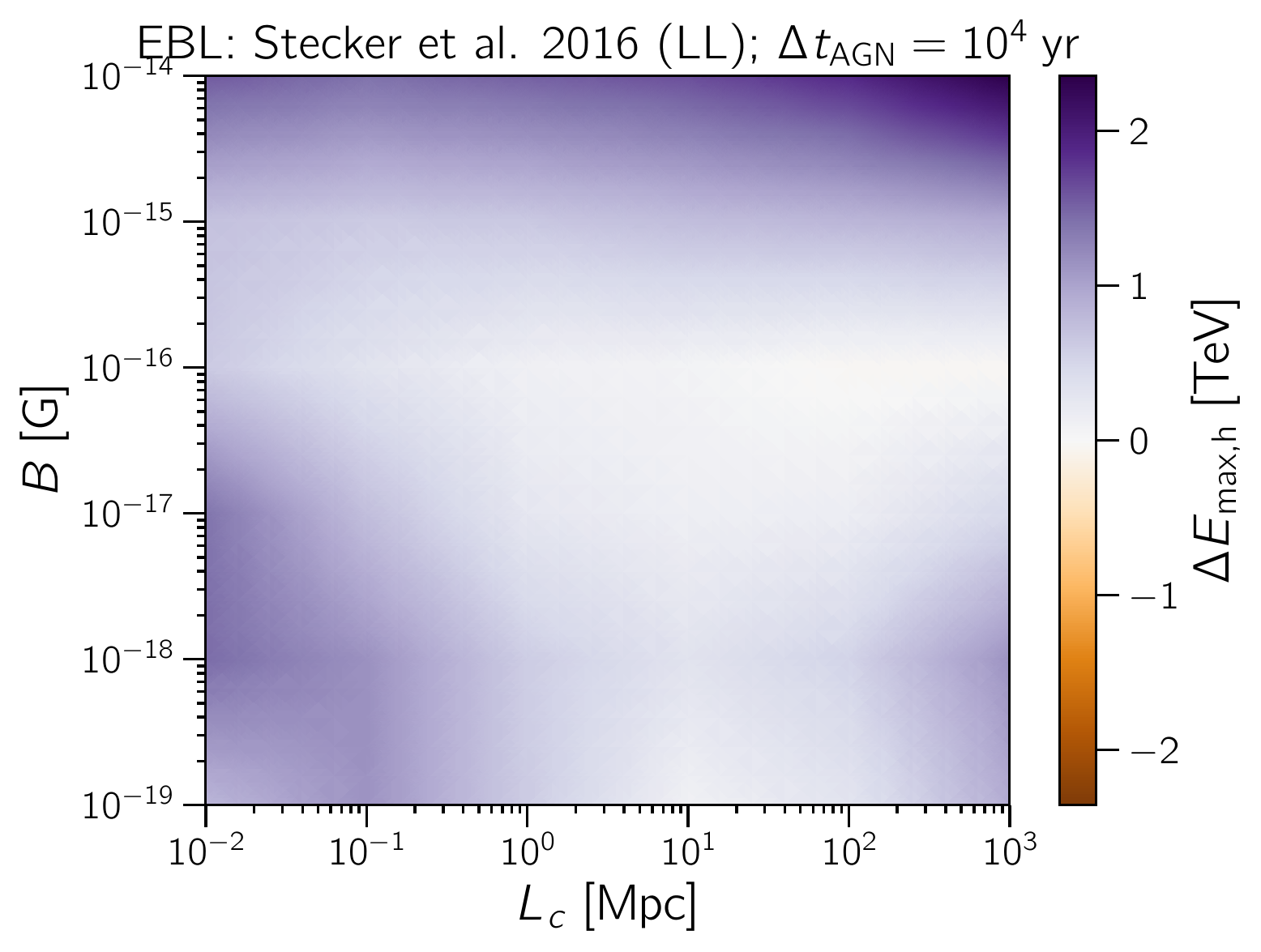}
  \includegraphics[width=0.49\textwidth]{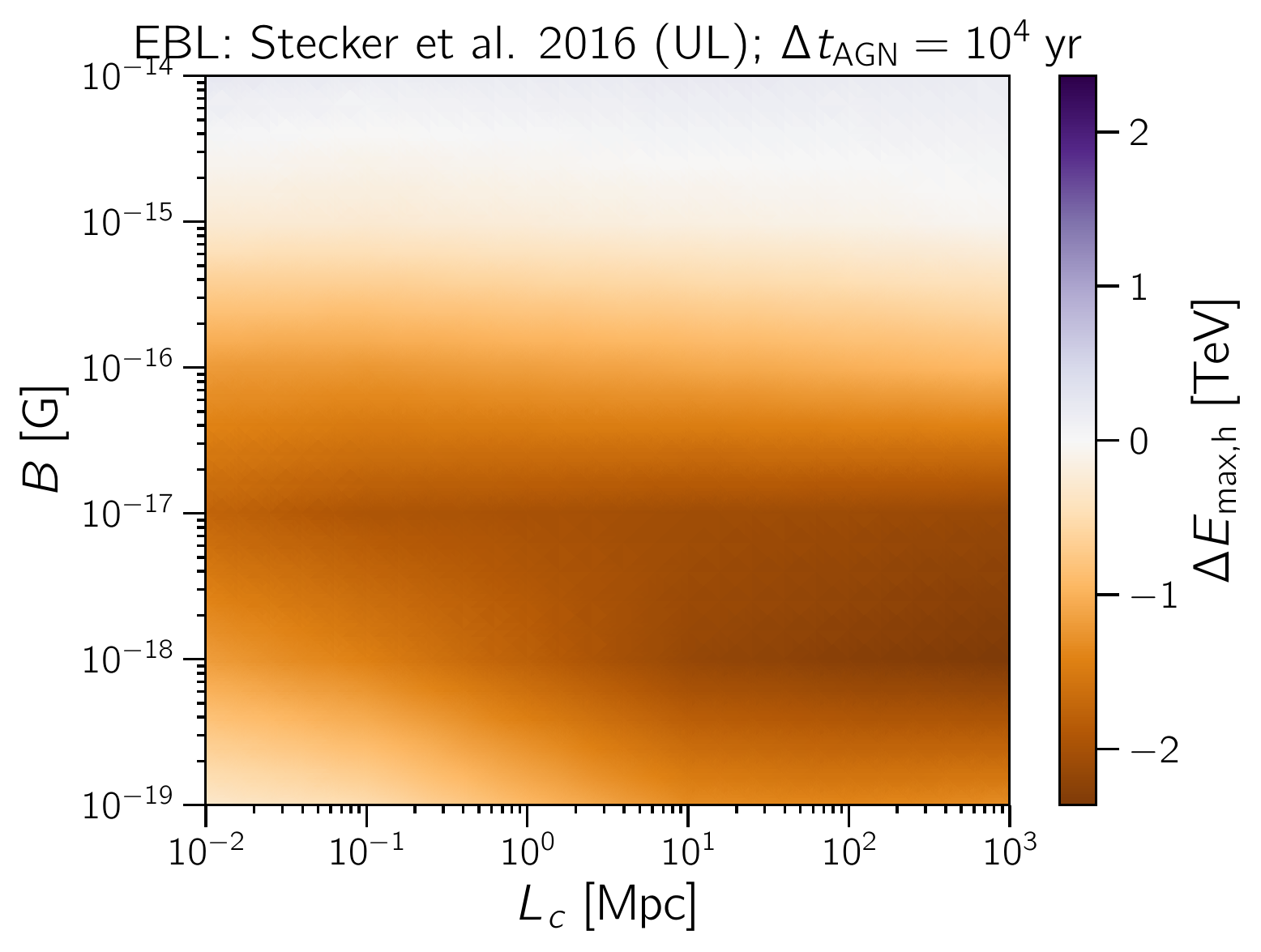}
  \caption{Average value of $\Emax$ (color scale) for different combinations of $B$ and $\Lc$. Here $\Delta \Emaxh$ denotes the difference between the best-fit $\Emaxh$ for a given pair $(B, \Lc)$ and the corresponding quantity for $B=0$. The panels correspond to the indicated EBL models, assuming that \txs{} is active over $\dtAgn =10^4 \; \text{yr}$ in the low state.}
  \label{fig:emax}
\end{figure}

With our statistical analysis we were able to analyze the influence of IGMFs on the inference of the intrinsic spectrum of gamma-ray sources~\cite{saveliev2021a}. For each magnetic field configuration described by $B$ and $\Lc$, there is a preferred source model, i.e.~a set of spectral parameters $(\alpha_{\rm l},E_{\rm max,l},\alpha_{\rm h},E_{\rm max,h},\eta$). The latter parameter set could, in principle, be correlated with the former. While the best-fit spectral indices ($\alphal$ and $\alphah$) are practically independent of the magnetic field configuration, we found that the same is not true for the maximum energy ($\Emaxl$ and $\Emaxh$). 
The change in the value of the maximal energy due to magnetic fields can be assessed by estimating the change in $\Emaxh$ with respect to the case without magnetic fields. We call this quantity $\Delta\Emaxh$. Its average value, marginalized over all other quantities except $B$ and $\Lc$, is shown in Fig.~\ref{fig:emax}. From this figure it is clear that the actual value of $\Emaxh$ may change considerably, depending on the EBL model, if magnetic field effects are considered. The two bottom panels, corresponding to the lower and upper limit EBL model by \cite{stecker2016a}, respectively, point to an interesting trend: In the presence of IGMFs, compared to the $B=0$ case, the best-fit value of $\Emaxh$ increases for \textit{all} considered magnetic field configurations in the case of the former EBL model, and decreases for \textit{all} combinations of $B$ and $L_{\rm c}$ for the latter one.

The detection of high-energy neutrinos correlating with the electromagnetic signal favors a hadronic or leptohadronic origin for the gamma-ray emission~\cite{magic2018a,keivani2018a,gao2019a} and strengthens the case for multi-TeV production in the blazar. The limits presented here are rather robust and rely only on the assumption that the gamma rays at $E \gtrsim 1 \; {\rm GeV}$ observed by Fermi-LAT and the highest-energy bin observed by MAGIC at $E \approx 400 \; {\rm GeV}$ are produced during the neutrino flaring period. Therefore, the reliability of our limits reflects the significance of this correlation. 

The multimessenger approach used to constrain IGMFs based on time delays between high-energy gamma rays and their counterparts (in neutrinos and other wavelengths) is powerful. It differs from the methods commonly found in the literature~\cite{plaga1995a,neronov2009a,neronov2013a} in that we do not attempt to correlate the~GeV and~TeV emissions with each other. Instead, we use the time delay between the neutrino and the gamma-ray signals. For this reason, this method enables us to probe the Universe up to high redshifts, since no~TeV signal is required to perform these estimates and we can evade the limitations placed by the EBL attenuation of the very-high-energy part of the spectrum.  

The idea presented in this work may be applied to other sources if the high-energy gamma-ray signal correlates with the arrival of high-energy neutrinos, since this guarantees the production of very energetic gamma rays \emph{inside} the source, even though there is no guarantee that they can escape the environment. If no neutrinos are observed, the time delays between the cascade photons and another messenger, such as gravitational waves (e.g. from mergers of compact objects), would rely on the existence of a putative multi-TeV emission which, in the case of high-redshift sources, cannot be observed. Either way, gamma rays in the GeV band, together with another messenger (neutrinos, gravitational waves, or TeV gamma rays) can be used to place limits on \emph{both} the strength \emph{and} the coherence length of IGMFs.

\section{Summary and Outlook} \label{sec:SummaryOutlook}

In this work we were able to show that flaring objects with simultaneous gamma-ray and neutrino emission may be used to place constraints on the properties of magnetic fields using the time delay between the signals of the two messengers. This is done through a rigorous statistical analysis of a wide range scan over both the spectral properties of the source and the IGMF-related parameters, namely its strength ($B$) and coherence length ($\Lc$) of the magnetic field. 

We have also shown that IGMFs may have a significant impact on the determination of intrinsic properties of gamma-ray sources. In particular, the maximum energy of gamma rays emitted by the source is usually derived by considering only interactions with the CMB and the EBL. We have shown, however, that, if magnetic fields are at play, they have to be taken into account as well.

In the future we will extend our analysis to other flaring objects to obtain more robust magnetic field limits. Furthermore, we are planning to extend the parameter space of the parameters considered, in particular considering higher magnetic field strengths.

\section*{Acknowledgements}

\noindent
The work of A.S. is supported by the Russian Science Foundation under grant no.~19-71-10018. R.A.B. is funded by the Radboud Excellence Initiative.


\begin{thebibliography}{99}

\bibitem{kulsrud2008a}
R.~M.~Kulsrud and E.~G.~Zweibel. 
\textit{Rep. Prog. Phys.} \textbf{71} (2008) 046901.

\bibitem{durrer2013a}
R.~Durrer and A.~Neronov. 
\textit{Astron. Astrophys. Rev.} \textbf{21} (2013) 62.

\bibitem{vachaspati2021a}
T.~Vachaspati. 
\textit{Rept. Prog. Phys.} \textbf{84} (2021) 074901.

\bibitem{neronov2009a}
A.~Neronov and D.~V.~Semikoz. 
\textit{Phys. Rev. D} \textbf{80} (2009) 123012.

\bibitem{jedamzik2019a}
K.~Jedamzik and A.~Saveliev. 
\textit{Phys. Rev. Lett.} \textbf{123} (2019) 021301.

\bibitem{alvesbatista2021a}
R.~{Alves Batista} and A.~Saveliev. 
arXiv:2105.12020 [astro-ph.HE] (2021).

\bibitem{neronov2010a}
A.~Neronov and I.~Vovk.
\textit{Science} \textbf{328} (2010) 73.

\bibitem{broderick2018a}
A.~E. Broderick \textit{et al.}
\textit{Astrophys. J.} \textbf{868} (2018) 87.

\bibitem{alvesbatista2019a}
R.~{Alves Batista} and A.~Saveliev.
\textit{Mon. Not. R. Astron. Soc.} \textbf{489} (2019) 3836.

\bibitem{alvesbatista2020a}
R.~{Alves Batista} and A.~Saveliev.
\textit{Astrophys. J. Lett.} \textbf{902} (2020) L11.

\bibitem{saveliev2021a}
A.~Saveliev and R.~{Alves Batista}.
\textit{Mon. Not. R. Astron. Soc.} \textbf{500} (2021) 2188.

\bibitem{icecube2018a}
IceCube Collaboration.
\textit{Science} \textbf{361} (2018) 147.

\bibitem{icecube2018b}
IceCube Collaboration \textit{et al.}
\textit{Science} \textbf{361} (2018) eaat1378.

\bibitem{paiano2018a}
S. Paiano \textit{et al.}
\textit{Astrophys. J. Lett.} \textbf{854} (2018) L32.

\bibitem{magic2018a}
MAGIC Collaboration.
\textit{Astrophys. J. Lett.} \textbf{863} (2018) L10.

\bibitem{parma2002a}
P. Parma \textit{et al.}
\textit{New Astron. Rev.} \textbf{46} (2002) 313.

\bibitem{alvesbatista2016a}
R. {Alves Batista} \textit{et al.}
\textit{J. Cosmol. Astropart. Phys.} \textbf{05} (2016) 038.

\bibitem{gilmore2012a}
R.~C. Gilmore \textit{et al.}
\textit{Mon. Not. R. Astron. Soc.} \textbf{422} (2012) 3189.

\bibitem{dominguez2011a}
A. {Dom{\'\i}nguez} \textit{et al.}
\textit{Mon. Not. R. Astron. Soc.} \textbf{410} (2011) 2556.

\bibitem{stecker2016a}
F.~W. Stecker \textit{et al.}
\textit{Astrophys. J.} \textbf{827} (2016) 6.

\bibitem{bertone2006a}
S. Bertone \textit{et al.}
\textit{Mon. Not. R. Astron. Soc.} \textbf{370} (2006) 319.

\bibitem{miniati2011a}
F. {Miniati} and A.~R. {Bell}
\textit{Astrophys. J.} \textbf{729} (2011) 73.

\bibitem{keivani2018a}
A. {Keivani} \textit{et al.}
\textit{Astrophys. J.} \textbf{864} (2018) 84.

\bibitem{gao2019a}
S. {Gao} \textit{et al.}
\textit{Nature Astron.} \textbf{3} (2019) 88.

\bibitem{plaga1995a}
R. Plaga.
\textit{Nature} \textbf{374} (1995) 430.

\bibitem{neronov2013a}
A. {Neronov} \textit{et al.}
\textit{Astron. Astrophys.} \textbf{554} (2013) A31.

\end{thebibliography}
\end{document}